\documentclass[particles,article,accept,pdftex,moreauthors]{Definitions/mdpi} 
\usepackage{subcaption}
\usepackage{diagbox}
\usepackage{siunitx}

\firstpage{1} 
\pubvolume{8}
\issuenum{2}
\articlenumber{48}
\pubyear{2025}
\copyrightyear{2025}
\externaleditor{Tommaso Dorigo, Roberto Ruiz de Austri Bazan, Jose Salt and Pietro Vischia }
\datereceived{31 January 2025} 
\daterevised{17 March 2025} 
\dateaccepted{19 April 2025} 
\datepublished{23 April 2025} 
\hreflink{https://doi.org/10.3390/\linebreak particles8020048} 

\Title{Development and Explainability of Models for Machine-Learning-\mbox{Based Reconstruction of Signals in Particle Detectors}}

\TitleCitation{Development and Explainability of Models for Machine-Learning-Based Reconstruction of Signals in Particle Detectors}

\Author{{Kalina} 
 Dimitrova *\orcidA{}, Venelin Kozhuharov \orcidB{} and Peicho Petkov \orcidC{}}

\AuthorNames{Kalina Dimitrova, Venelin Kozhuharov and Peicho Petkov}

\AuthorCitation{{Dimitrova}
, K.; Kozhuharov, V.; Petkov, P.}

\address[1]{%
Faculty of Physics, Sofia University ``St. Kliment Ohridski'', 5 J. Bourchier Blvd., 1164 Sofia, Bulgaria; {venelin@phys.uni-sofia.bg (V.K.); peicho@phys.uni-sofia.bg (P.P.)} 
}

\corres{\hangafter=1 \hangindent=1.05em \hspace{-0.82em}Correspondence: kalina@phys.uni-sofia.bg}

\abstract{Machine learning methods are being introduced at all stages of data reconstruction and analysis in various high-energy physics experiments. We present the development and application of 
convolutional neural networks with modified autoencoder architecture
for the
reconstruction  of 
the pulse arrival time and amplitude in individual scintillating crystals in 
electromagnetic calorimeters and other detectors.
The network performance is discussed as well as the application of xAI methods
for further investigation of the algorithm and improvement of the output accuracy.}


\begin{document}


\section{Introduction}
One of the main steps in high-energy physics data analysis is event reconstruction. It deals with the task of transforming the detector output into information about the particles that have entered the detector. When photons or charged particles enter segmented scintillating crystal calorimeters, they create electromagnetic showers which result in many pulses, recorded by the readout system of the detector. The correct reconstruction of the arrival times and amplitudes of these pulses is the first step towards the accurate determination of the particle's arrival time and energy, and, eventually, the reconstruction of entire physics events.

Signal reconstruction becomes a complicated task in cases where high particle multiplicities lead to 
a
large number of pulses in short time periods. Different methods are being developed in an attempt to deal with this problem, including 3D clustering~\cite{bib:3Dclu} and pulse shape discrimination~\cite{bib:PSD}. Machine learning methods find growing application for close-in-time pulse separation in many high-energy physics experiments~\cite{bib:aad}. 

Explainable AI (xAI) techniques aim to provide better understanding of how the output of complex machine learning models is generated. Their application is not only important for explaining the process of obtaining the results and bringing insights into which components of the input data are the most important for formulating the final result, but may also point towards possible improvements that can be introduced to the algorithms. Most of the common xAI techniques like Integrated Gradients~\cite{bib:intgrad}, SmoothGrad~\cite{Smilkov2017}, Occlusion Sensitivity~\cite{Valois2023}, and others were initially developed for studying the performance of image classification models. These methods are based on introducing changes to the input data (scaling the values for the Integrated Gradients, adding noise for SmoothGrad, or masking regions for Occlusion Sensitivity) and monitoring the output of the network for the altered data. Saliency maps are constructed on the basis of the results, highlighting which regions of the images bring the biggest change in the output when altered. These maps show the regions that have the highest importance for the image classification. Explainability methods have successfully been implemented for understanding one-dimensional models for tasks like time series data analysis~\cite{bib:rojat}.

In this work, we present the use of neural networks with modified autoencoder (MAC) architectures for pulse arrival time and amplitude determination. Different versions of the Occlusion Sensitivity method were applied to one of the trained models and the results were used for explaining the arrival time determination.
In addition, those results imply that the sensitivity can be further improved.
A new upsampling version of the modified autoencoder (UMAC) was developed. It provides better time resolution which is an important factor for the multiple pulse separation ability of the models.

The TensorFlow~\cite{bib:tf} and Keras~\cite{bib:keras} libraries are used for developing the models and the xAI investigations. The results are visualized using the matplotlib~\cite{bib:matplotlib} package and the ROOT~\cite{bib:ROOT} framework. The neural network architectures are visualized using the visualkeras~\cite{bib:viskeras} package.

\section{Convolutional Neural Networks for Signal Reconstruction}
Simple classifying neural networks with convolutional and dense layers can be used for determining the number of pulses in the waveforms, produced in individual channels of particle detectors~\cite{bib:calor}.
The successful recognition of individual signals prompts further investigation of the application of neural networks for denoising the waveforms and extracting signal parameters~\cite{bib:nafski22}, and 
convolutional autoencoders are successfully applied for such signal processing tasks~\cite{bib:azuara}.
The models used in this study have convolutional autoencoder
architectures starting with an encoder with three convolutional layers.
The first one consists of 128 filters with kernel size 18, the second one has 64 filters with kernel size 12, and the third one has 32 filters with kernel size 6. Two dropout layers with rate 0.2 are placed between them.
The encoder is followed by a decoder of three deconvolutional layers, mirroring the filter numbers and kernel sizes of the encoder. The first transposed convolution layer in it has 32 filters with kernel size 6, the second has 64 filters with kernel size 12, and the third has 128 filters with kernel size 18. Again, two dropout layers for regularization purposes with 0.2 rate are placed between the transposed convolution layers of the decoder. 
All convolution and deconvolution layers in the encoder and decoder use ReLU activation.
The output layer is a transposed convolution layer with 1 filter with kernel size 18 and no activation function. 
Applying such a model to the input waveform results in a version of it in the output with the noise reduced. In contrast, in this study, such architecture is used for reconstructing the arrival time and amplitude of the pulse this waveform represents. Namely, the output preserves the dimensions of the input but it only contains the amplitude values at the pulse arrival time positions. 

\begin{figure}[H]
\centering
\includegraphics[width=0.49\columnwidth]{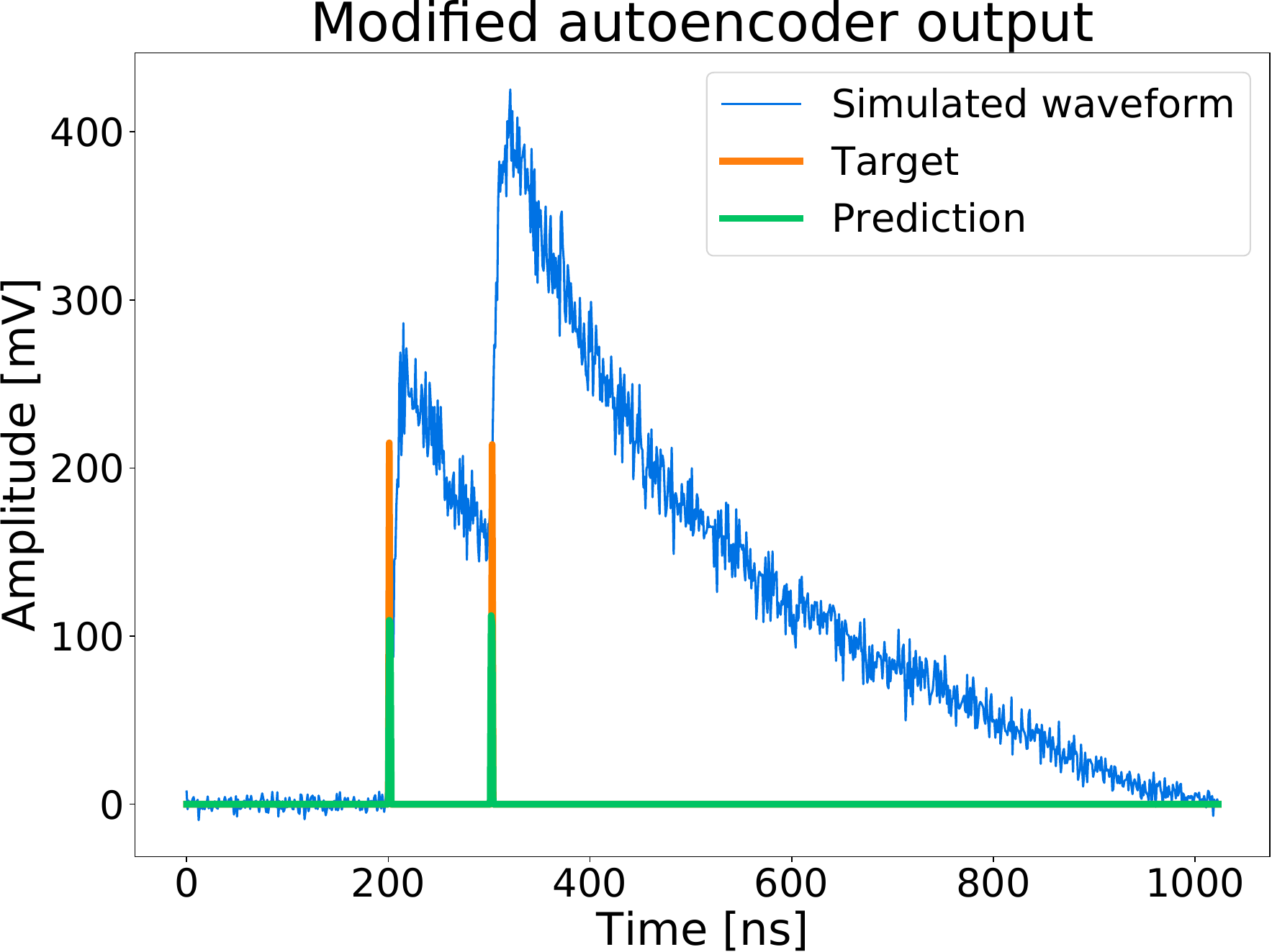}
\includegraphics[width=0.49\columnwidth]{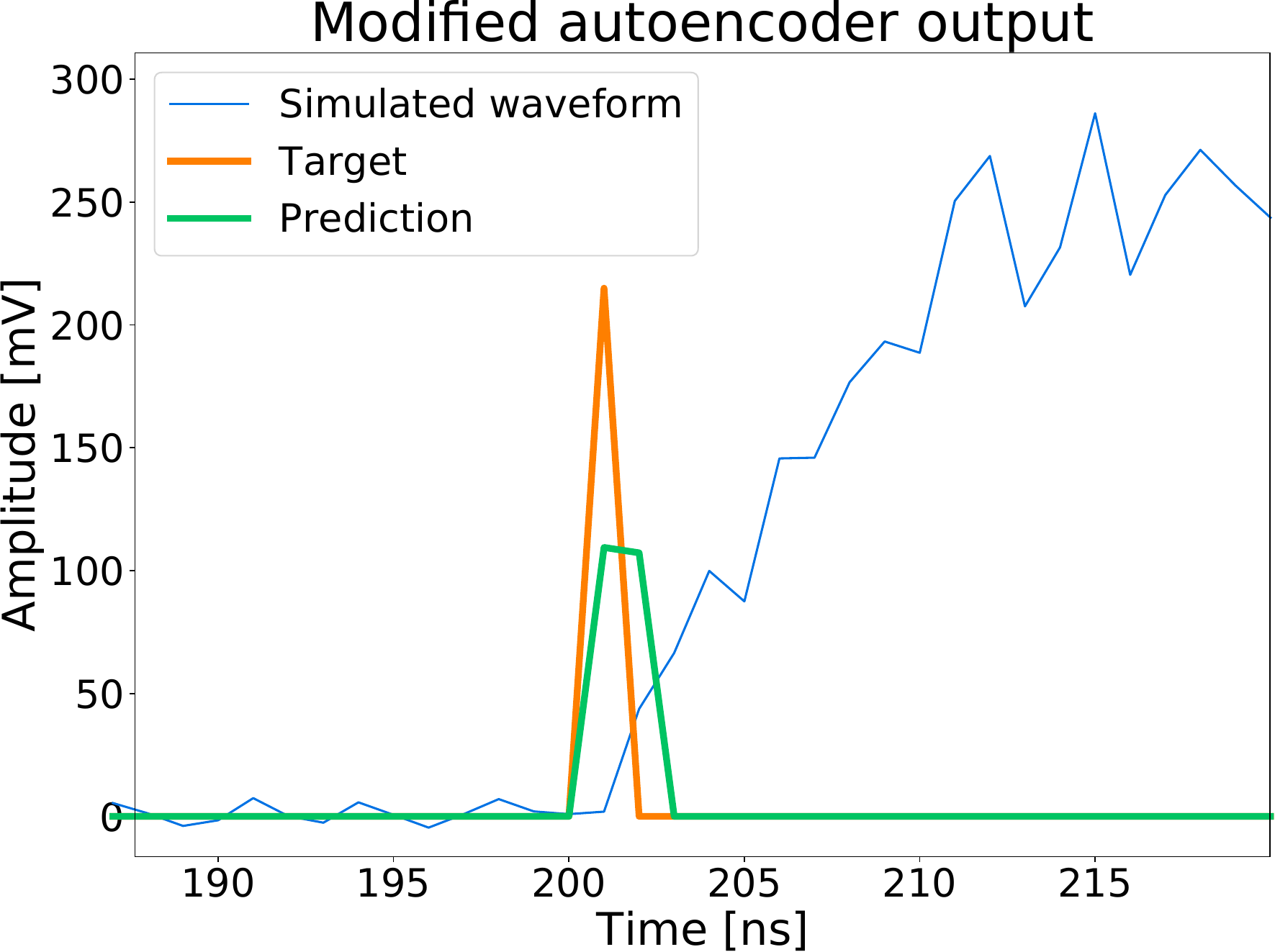}
\caption{An example of a simulated event, used for training the MAC model. (\textbf{Left}) The simulated waveform (blue), used as input to the model and the model output (green), which aims to replicate the label (orange), used for this event. (\textbf{Right}) A detailed look into the region around the arrival of the first pulse from the same event with the simulated waveform (blue), the true desired output value (orange), and the actual model output (green)~\cite{bib:nafski22}.
\label{fig:autoencoders}}
\end{figure}

A training dataset for such models contains the input waveform arrays and their corresponding desired output arrays, referred to as labels in the text below.
The labels have the same shape as the input arrays for each event, consisting of zero values except for the pulse arrival time positions, where the signal amplitude value is placed. This label 
carries all the information that needs to be extracted from the waveforms.
The input to the model remains the raw waveform, produced in the simulation. The activation of the output layer is also set to ReLU.
An example waveform with the model output is provided on the left panel of Figure~\ref{fig:autoencoders}. 
The output from applying the models gives several non-zero values on the neighboring positions next to the main one, as shown on the right panel of Figure~\ref{fig:autoencoders}. 
During the result post-processing, a merging window $T$ is defined and all values on positions $t_i$ for which $|t_{max}-t_i|<\frac{T}{2}$ are merged into the value on the position of the maximum $t_{max}$, which is declared to be the arrival position.

The training data were chosen to match the form of the output of the 
CAEN V17XX (e.g., CAEN V1742, V1751, etc. ~\cite{bib:caen})
family of digitizers. 
They are often used  in test setups and 
in 
a few 
high-energy physics experiments to digitize the input signals.
The investigated model was trained using simulated datasets of 
events with 1024 values length, with each value representing a time interval of 1~ns. 
This length was chosen to match the length of the events, produced by V1742 boards that use chips with 1024 storage capacitors. 
For V1751, the length can be adjusted, but 1024 is a common choice. 
The events contain a maximum of five pulses with a shape defined as 
\begin{equation}
    A(t)=A_0~(e^{\frac{-(t-t_0)}{\tau_1}}-e^{\frac{-(t-t_0)}{\tau_2}})=A_0~e^{\frac{-(t-t_0)}{\tau_1}}(1-e^{-(t-t_0)(\frac{1}{\tau_2}-\frac{1}{\tau_1})}), ~~~~t\geq t_0,
\end{equation}
where $t_0 $ is the pulse arrival time, 
$\tau_1$ and $\tau_2$ are the signal fall and rise times, 
and $A_0$ is the pulse amplitude. The signal fall $\tau_1$ is determined by the type of scintillator used in the detector, while the signal rise $\tau_2$ comes from the specifics of the electronics, used in the system. The value for the amplitude $A_0$ has to be within the range 
of the V17XX family of digitizers, which is typically
1~V.
The pulse arrival times follow a uniform distribution and are generated with 
much higher precision ($10^{-6}$~ns), 
so the final arrival time values are rounded to 1~ns. Testing the model on independent datasets shows 
a time resolution of $\sim$0.5~ns.

One of the models trained on simulated data was introduced to the reconstruction 
of real data from the electromagnetic calorimeter of a fixed-target experiment~\cite{bib:nafski23}.
The reconstructed data were analyzed by performing a selection of annihilation events in
order to evaluate their characteristics obtained by the ML model and compare them to the conventional reconstruction algorithm, employed by the experiment, which relies on peak finding.
The results show that the ML model needs additional calibration in order to match the particle energy. The reconstruction algorithms it is compared to have also been calibrated in a similar manner. 
The application of the ML model, however, shows a higher number of low-energy events recognized than the peak finding method, as well as a better time resolution.

\section{Application of the Occlusion Sensitivity xAI Method}

Understanding which regions of the individual pulses are the most important for distinguishing them is a necessary step for 
understanding the critical signal parameters.
Different xAI methods can be applied to the input data for this task~\cite{bib:varna}, of which Occlusion Sensitivity provides the most insight into the model sensitivity in this case.
In image processing, the Occlusion Sensitivity method 
shows whether models actually locate the relevant objects when classifying images or, rather, learn from some patterns in their surroundings~\cite{bib:occ}. This is achieved by occluding different regions of the input image with a gray or black mask and monitoring the model output for these modified inputs.

In the case of particle signals in calorimeters, we define a mask of fixed length of 18~values set to zero and move it along the waveform. The model is applied to the input event for each consecutive position of the mask, and the value of the loss function is calculated for the prediction. The MAC in this study uses a mean square error loss function:

\begin{equation}
    \text{MSE} = \frac{1}{N} \sum_{i=1}^{N} \sum_{j=1}^{M}(y_j - \hat{y}_j)^2_i
\end{equation}
where 
$N$ is the number of samples in the dataset ($N=10^6$ for the investigated model), 
$M$ is the event length, 
$y_j$ is the true value on the $j$-th position in the event, 
$\hat{y}_i$ is the predicted value for the $j$-th position in the event,
and $i$ is the event index.


For a single event that is being predicted, this is reduced simply to the square error
\begin{equation}
    \text{SE} = \sum_{j=1}^{M} (y_j - \hat{y}_j)^2.
\end{equation}

\begin{figure}[H]
\includegraphics[width=0.47\columnwidth]{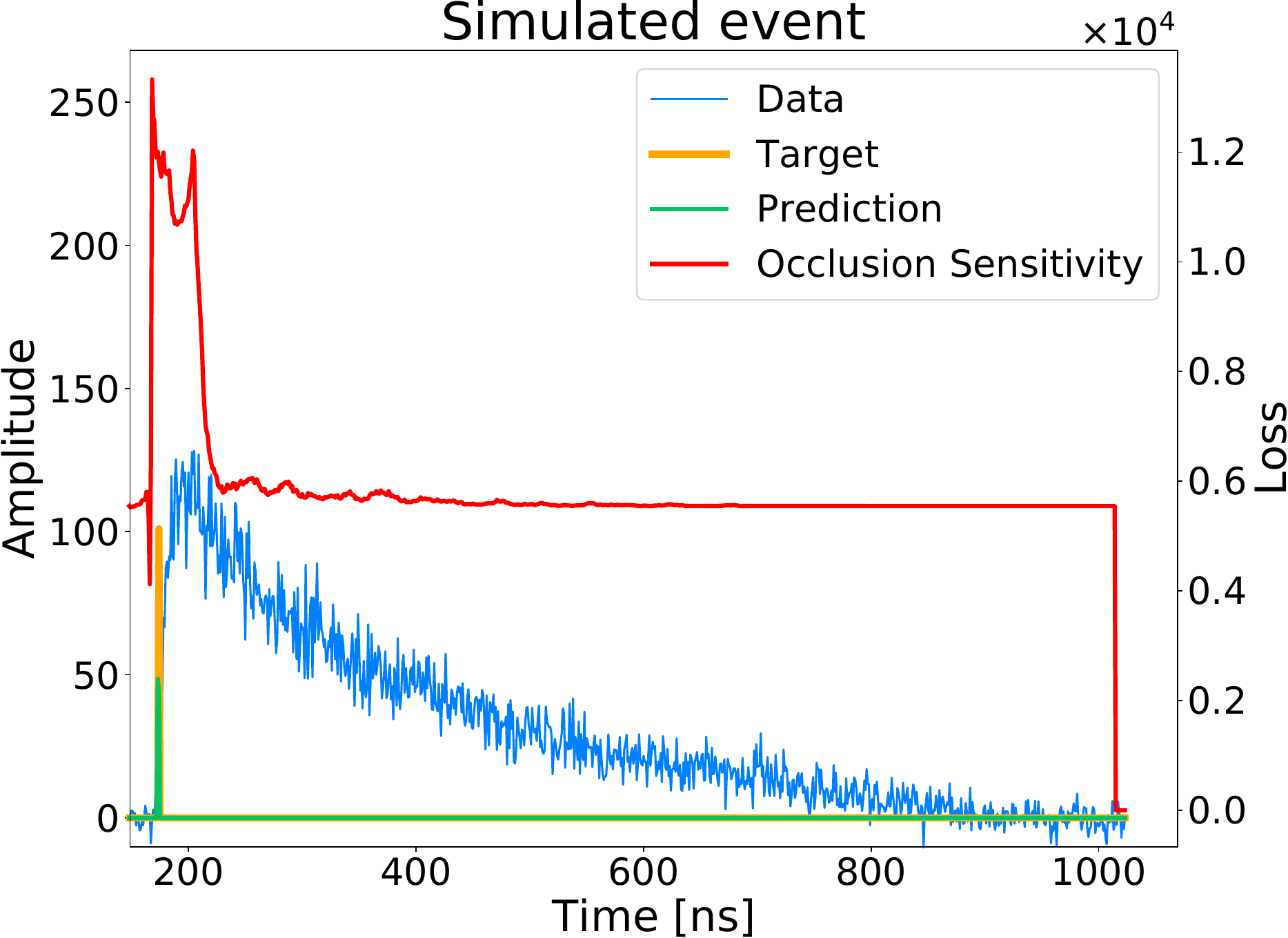}
~
\includegraphics[width=0.48\columnwidth]{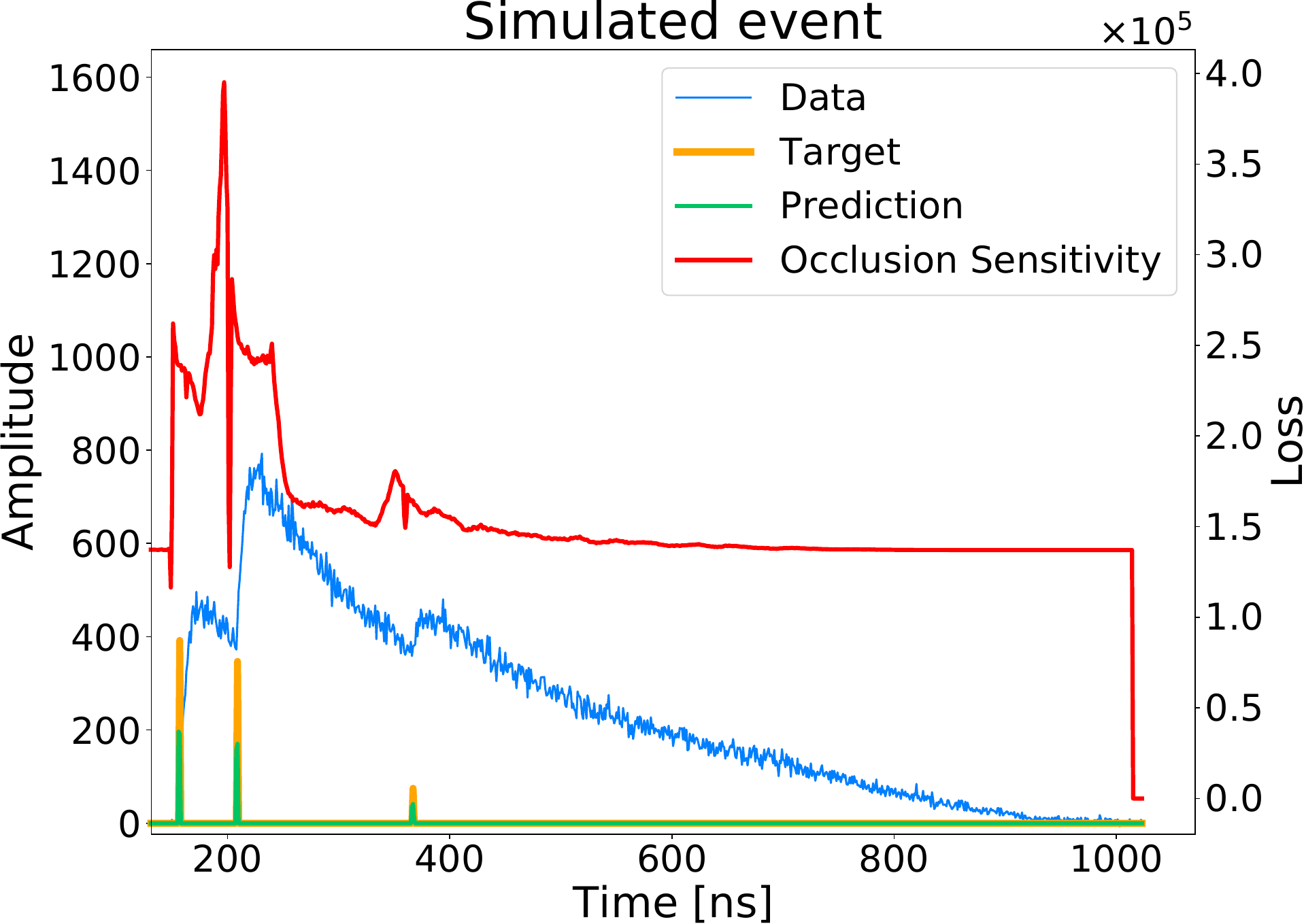}
\caption{Output of the Occlusion Sensitivity method for two events from the testing dataset. (\textbf{Left})~An event with a single pulse. The region of the signal front and maximum have the highest influence on the loss. (\textbf{Right}) An event with three pulses. 
Similarly, the biggest changes in the loss are observed at the fronts of the individual pulses.
\label{fig:occ}}
\end{figure}

\begin{figure}[H]
\centering
\includegraphics[width=0.49\columnwidth]{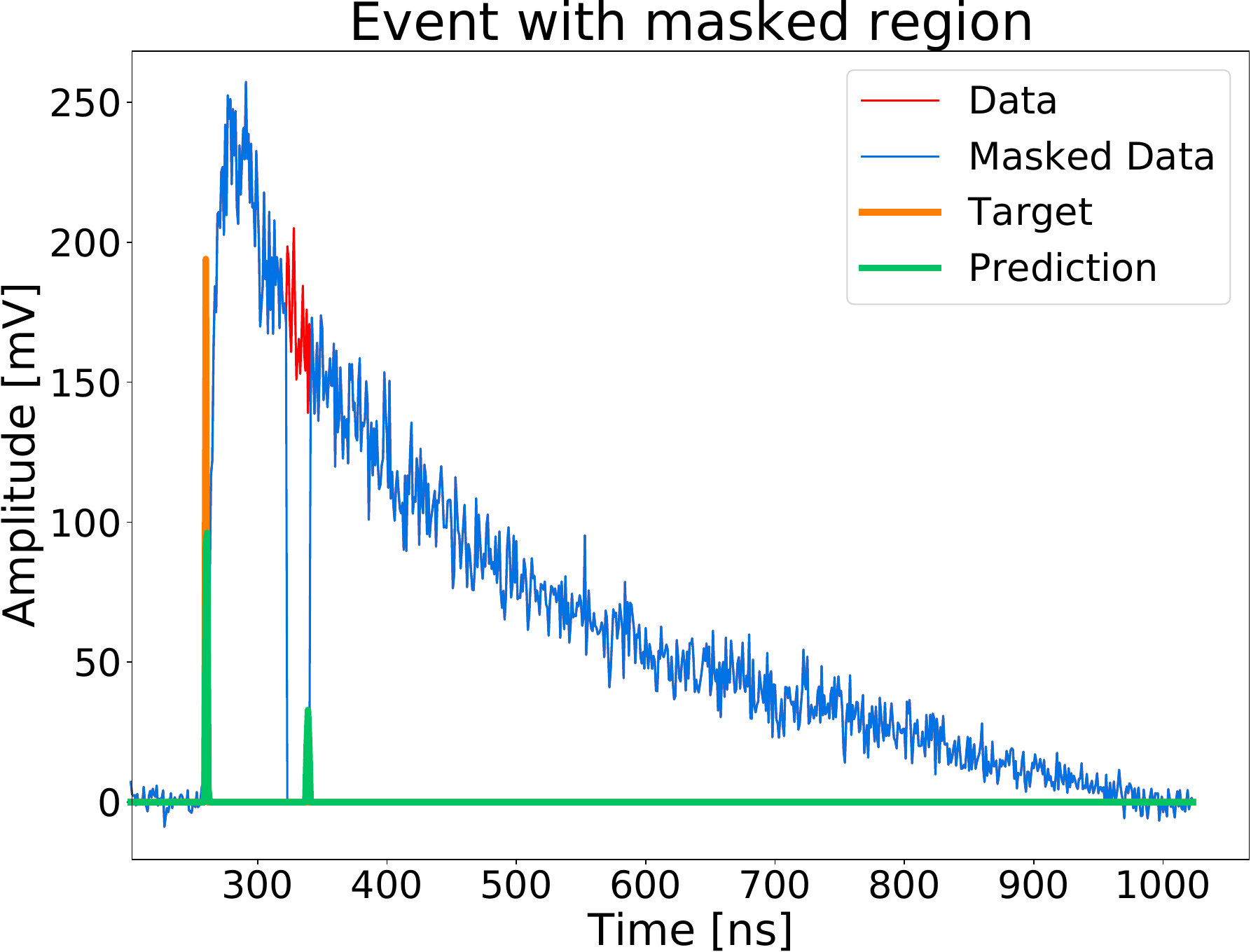}
\includegraphics[width=0.49\columnwidth]{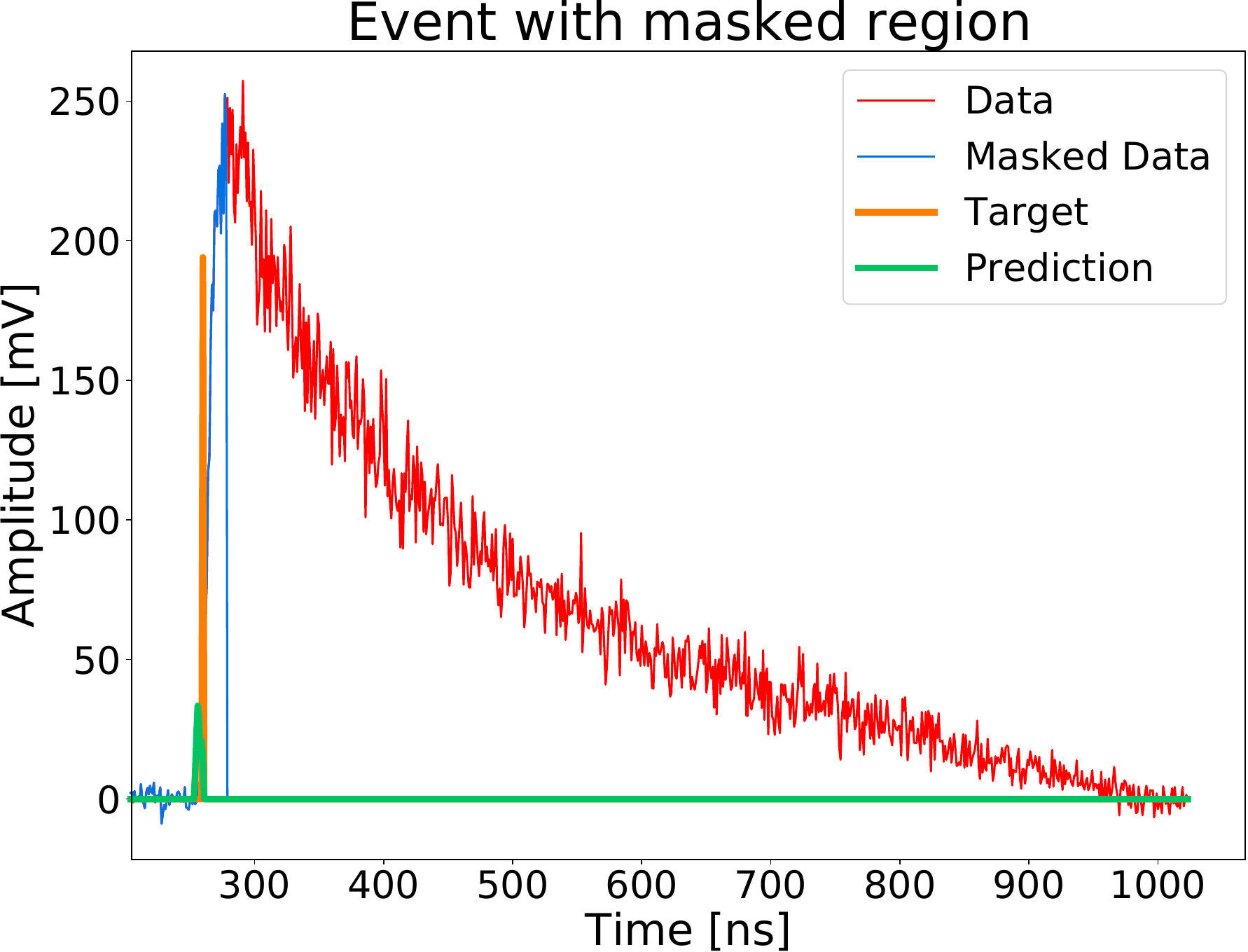}
\caption{An event, masked in two different ways for Occlusion Sensitivity investigations. (\textbf{Left})~A mask of size 18 is moved along the waveform and the loss is calculated for each position of the mask center. The example shows the mask, placed in the signal fall region. The calculated loss for the data with a masked region  is displayed in blue. The red portion of the waveform is set to 0. A fake prediction is made by the model where the mask ends. (\textbf{Right}) The whole event is occluded and values are revealed one by one. The example shows only the signal rise and a few values from its maximum region revealed. A prediction is made by the model; however, its value is much smaller.
\label{fig:masked}}
\end{figure}

Figure ~\ref{fig:occ} shows the results from applying Occlusion Sensitivity to two events with a different number of pulses in them
by monitoring the loss depending on the mask center position. An example of a masked event can be seen on the left panel of Figure~\ref{fig:masked}.
The case of a single pulse shows a rapid rise in the total loss when the signal rise region is occluded, followed by a period of relatively high loss values around the signal maximum. The loss quickly returns to a base level after the maximum region of the signal. This highlights the signal rise as one of the parts of the waveform, important for its recognition. 
Such behavior can be expected, as the signal beginning is where the non-zero value in the labels is. 
When making predictions on occluded waveforms, the model identifies fake pulses where the mask ends and the waveform is restored. 
Depending on the position of the mask along the waveform, this fake pulse would have a different height; hence, it provides smaller contribution to the total loss the further from the maximum the mask is.
However, revealing the whole signal rise and occluding the region of maximum values after it results in a secondary spike of high values of the total loss. This shows that not only is the steepest part of the waveform---the signal rise---important for identifying it, but also the relatively flat region around the peak. 

If an event contains multiple pulses, a similar pattern can be observed for each individual one.

\begin{figure}[H]
\hspace{-8pt}\includegraphics[width=0.75\columnwidth]{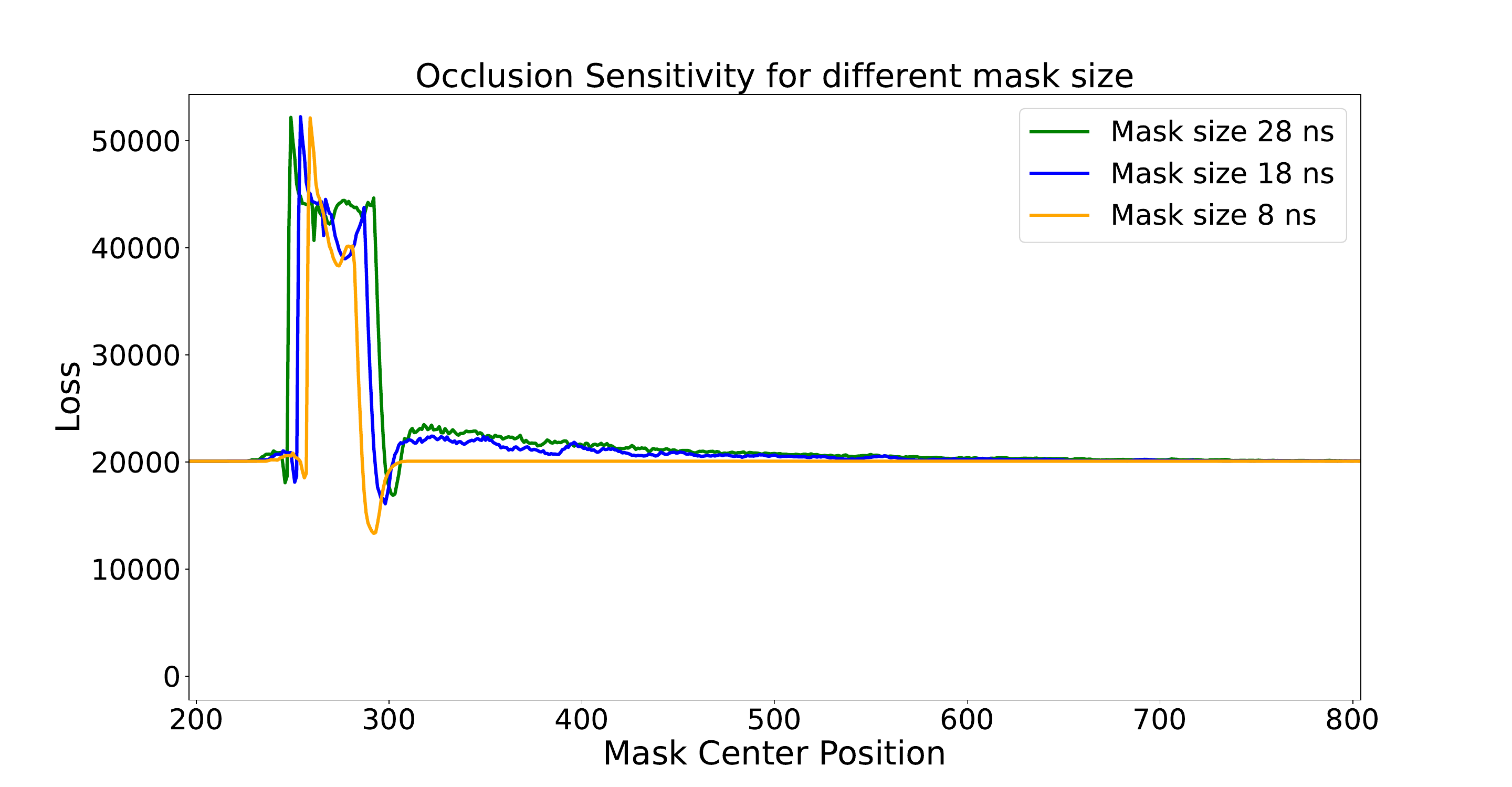}
\caption{{Comparison} 
 of the output of the Occlusion Sensitivity method for different mask sizes on the same event. The same pattern is observed; however, larger mask sizes provide more detail in the total loss change.
\label{fig:masksize}}
\end{figure}

The mask size has little influence over the overall shape of the Occlusion Sensitivity output. This can be seen in Figure~\ref{fig:masksize}, where the output is compared for three different lengths of the mask. The pattern remains similar; however, larger mask sizes result in more detailed output.

\vspace{-12pt}

\begin{figure}[H]
\includegraphics[width=0.49\columnwidth]{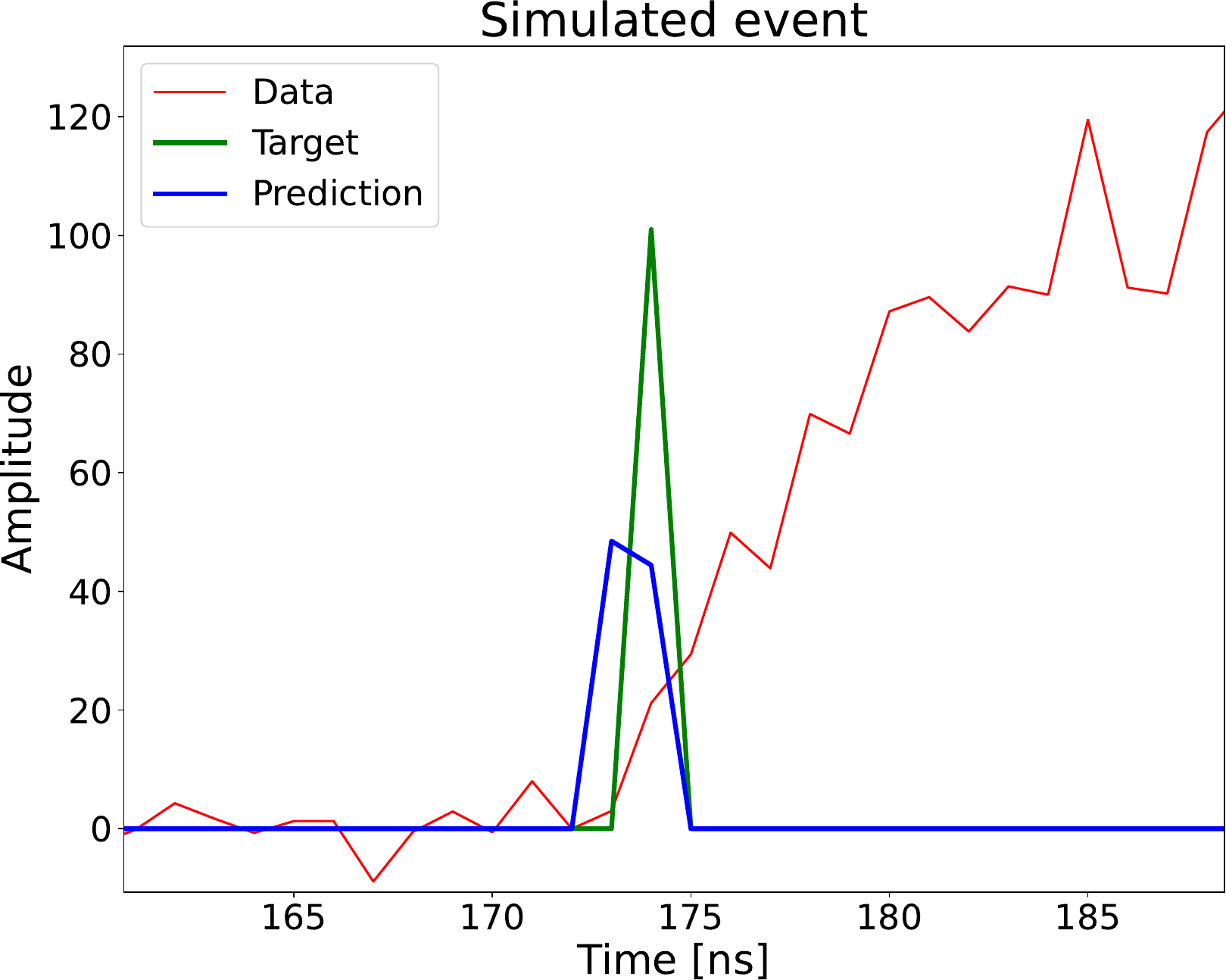}
\includegraphics[width=0.49\columnwidth]{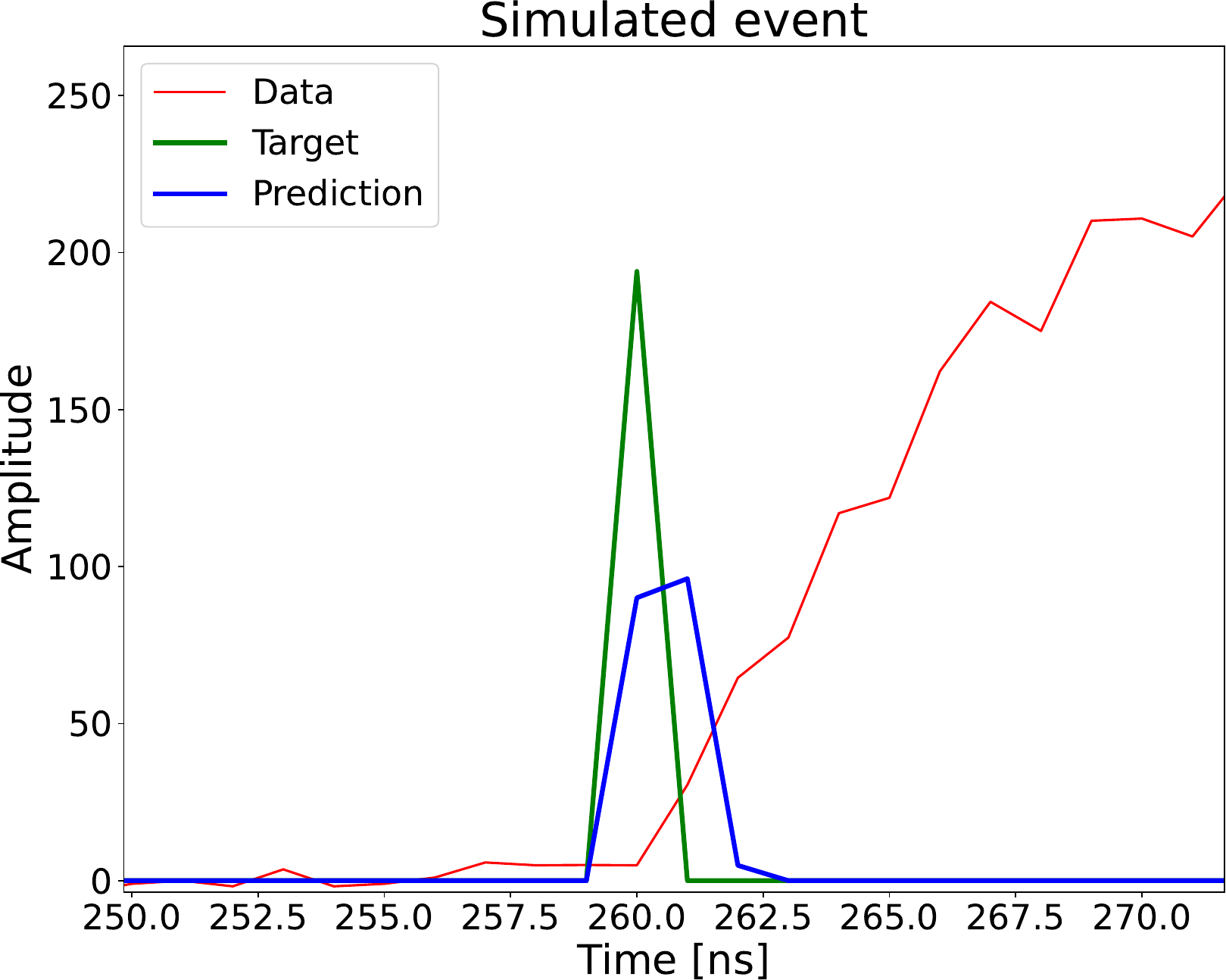}
\par\bigskip
\includegraphics[width=0.49\columnwidth]{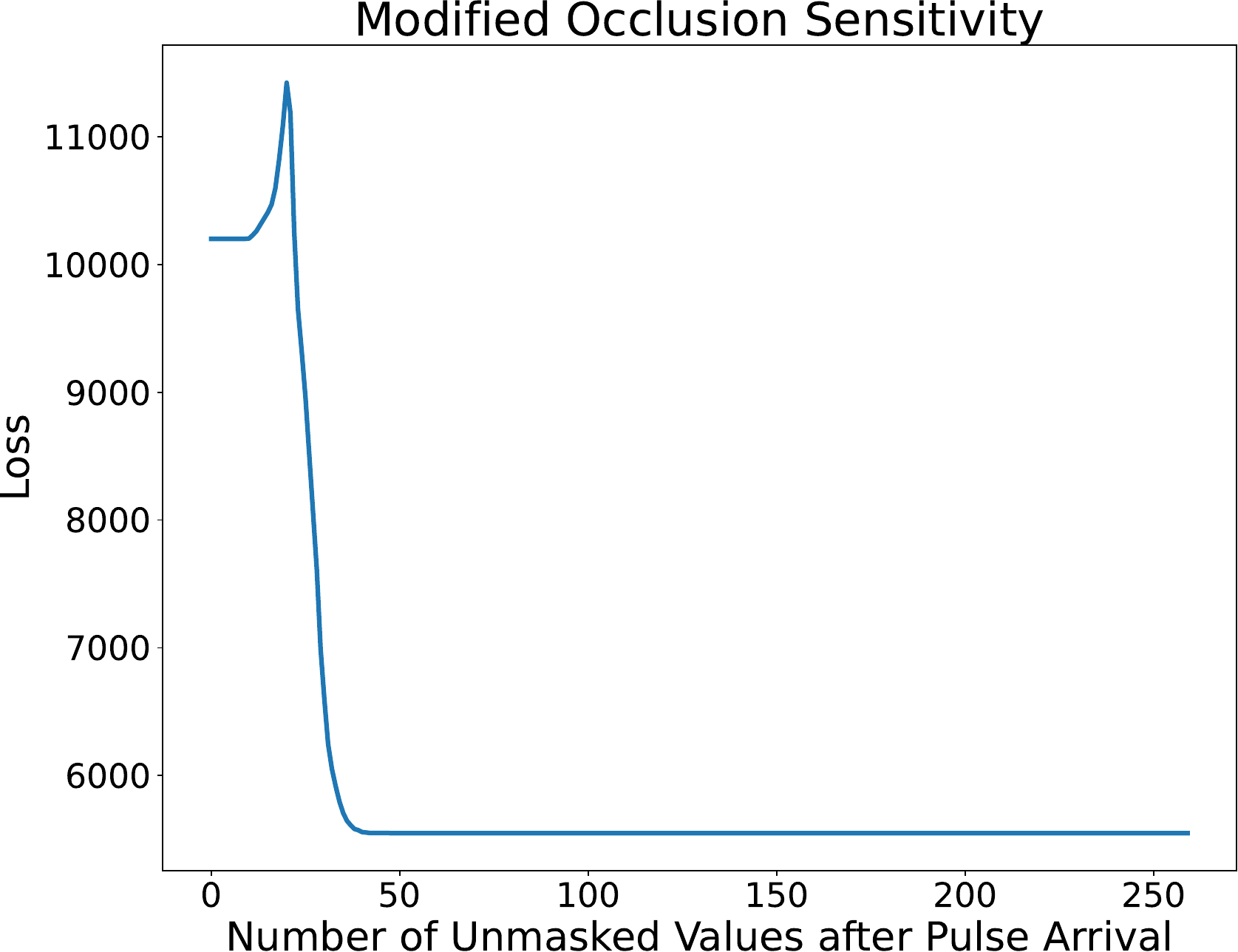}
\includegraphics[width=0.49\columnwidth]{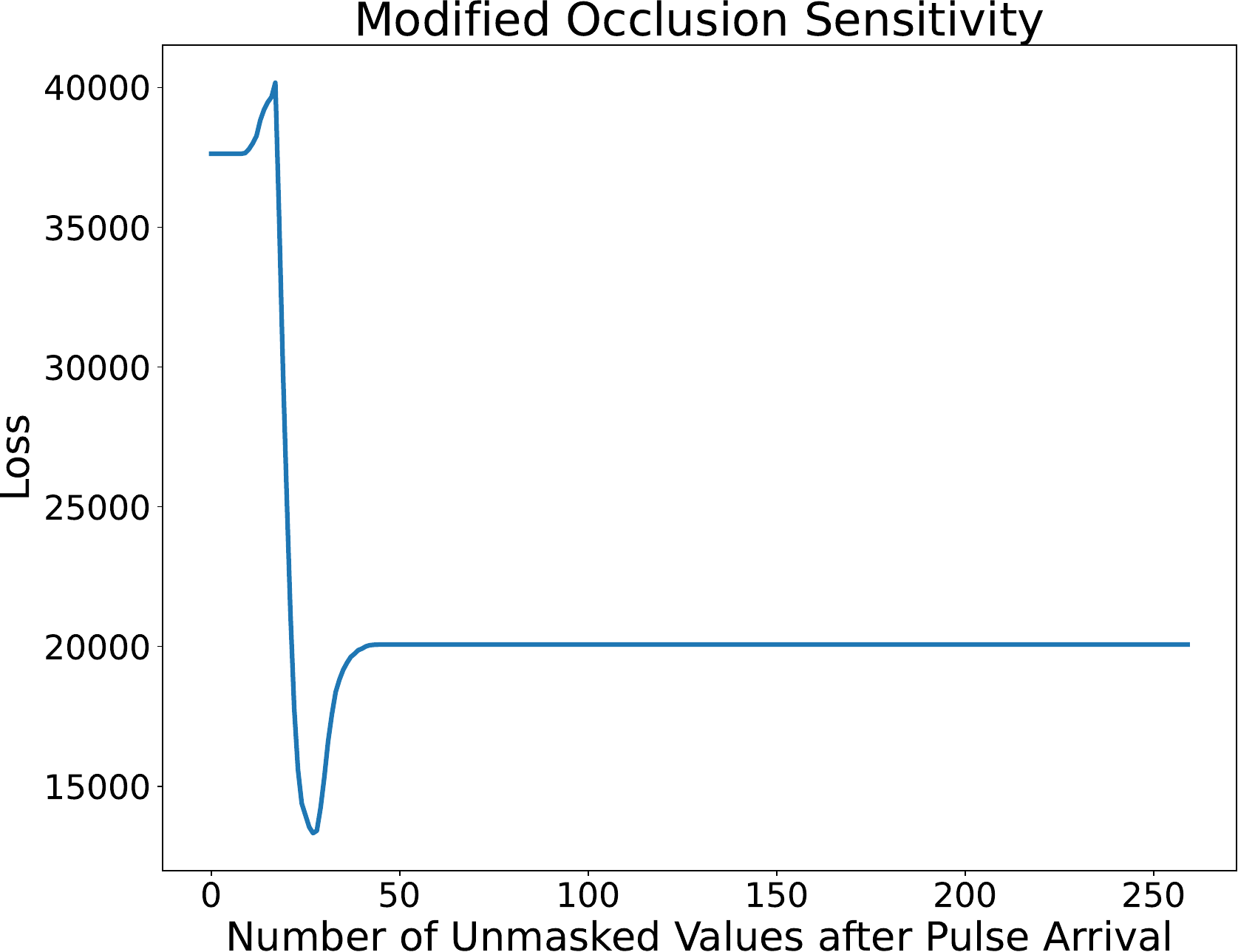}
\caption{{Output} 
 of the modified Occlusion Sensitivity for two events, each containing a single pulse. (\textbf{Left}) An event where the prediction is placed earlier than the rounded value position of the true arrival time. The loss remains constant at its lowest value after all the relevant for the recognition parts of the signal are unmasked. (\textbf{Right}) An event in which the prediction is placed after the rounded value position of the true arrival time. After reaching a minimum, the loss returns to a higher value before remaining constant.\label{fig:modocc}}
\end{figure}

A more in-depth investigation of the most important regions of the signal waveforms was performed by applying a modified version of Occlusion Sensitivity only to events with 
a single pulse.
Initially, the entire event is occluded, after which the values are revealed 
consecutively,
with the model being applied and the loss calculated.
An example of an event occluded in such a manner can be seen on the right panel of Figure~\ref{fig:masked}. It shows the stage where the whole signal rise and part of the maximum region are revealed, and the rest of the event is set to zero. The prediction shows a spike where the signal beginning is; however, the revealed part is not enough for a full prediction to be made.
Figure~\ref{fig:modocc} shows the total loss as a function of the number of unmasked values after the signal arrival. Two main possibilities for this dependence are observed for the different events: after reaching a minimum, the loss value remains constant at this value until the end of the event or rises to a higher value and then remains at it.

The difference in the modified Occlusion Sensitivity output can be attributed to the relationship between 
the label value for the signal arrival and the prediction. The arrival in the label is placed on the position, corresponding to the rounded value due to the binning signal arrival time from the simulation. 
If the predicted signal arrival is placed before this true arrival position, the loss remains at the minimal value. If it is placed at the same or a later position, the loss rises and remains at a higher value.


\section{Development and Evaluation of Upsampling Models}
\begin{figure}[H]
\includegraphics[width=0.75\columnwidth]{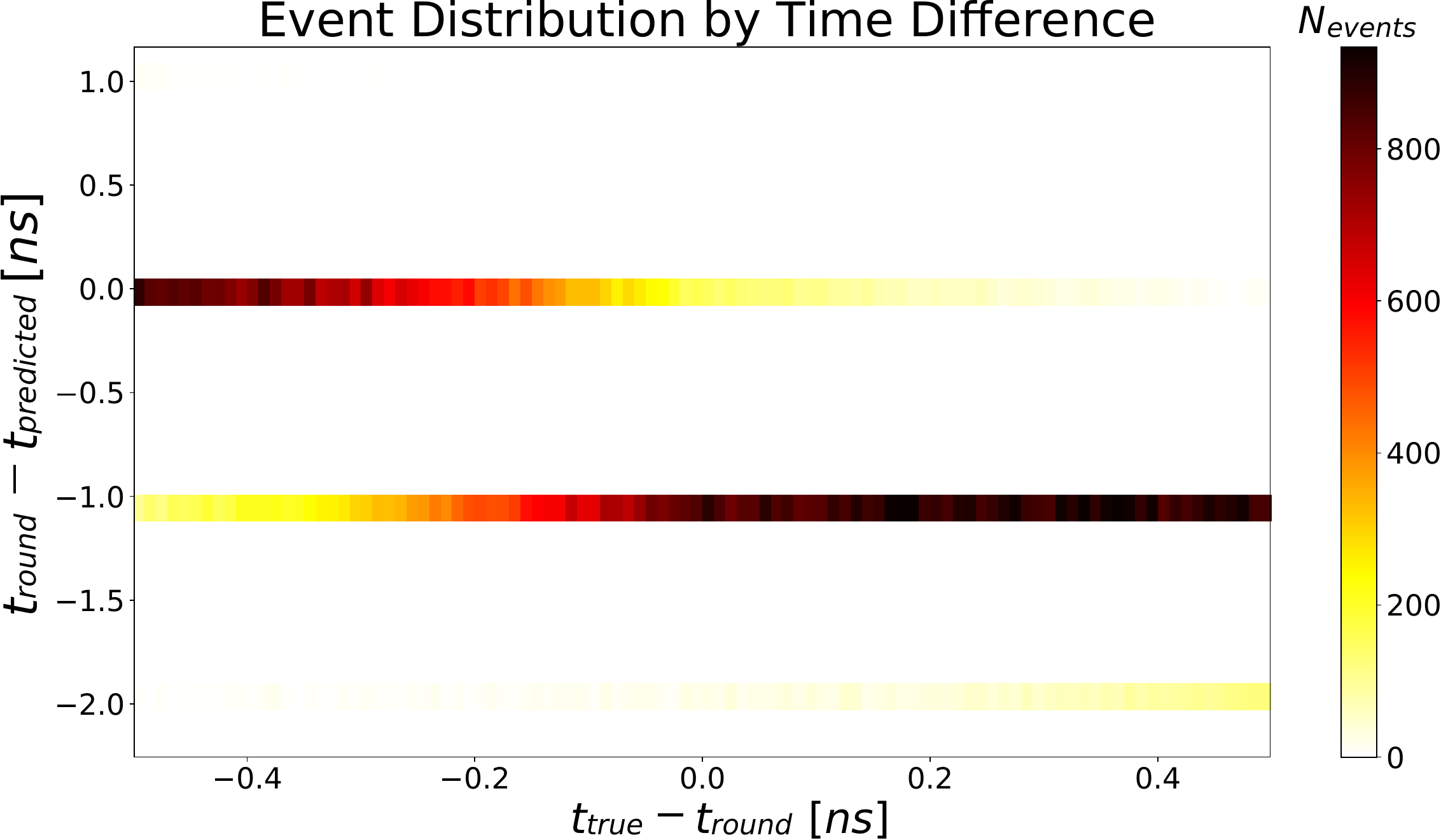}
\caption{Distribution of the 
difference between the true and rounded values in the label arrival times ($t_{true}-t_{round}$) on the x axis and the difference between the predicted and rounded values in the label arrival time ($t_{round}-t_{predicted}$) on the y axis. Events that actually arrive before the rounded position are also predicted earlier, while ones that arrive later are also predicted later.\label{fig:histo}}
\end{figure}

The results from applying the Occlusion Sensitivity method suggest that although the labels used when training the model use rounded values of the arrival time, there are two distinct patterns of the loss which are connected to the position of the prediction relative to the position in the label. 
Comparing the predictions to the exact arrival times shows that if the true arrival time is smaller than its rounded value in the label, the maximum in the prediction is also earlier. If the true arrival time is later than its rounded value, the maximum in the prediction is also later. 
This comparison indicates that the model is sensitive to which part of this time bin the signal truly begins in, and determining the arrival time for the different events by the model can be achieved with higher precision than the time bin width.
Figure~\ref{fig:histo} illustrates this comparison by showing the distribution of the events 
as a function of the difference between the true time $t_{true}$ and time bin position $t_{round}$ and the difference between the true time $t_{true}$ and the prediction $t_{predicted}$. Most of the events that have a true arrival time smaller than the time bin position also have earlier arrivals in the predictions; similarly, events with later true arrivals are also predicted later than the time bin position.

\vspace{-6pt}

This result 
prompts a change in the post-processing algorithm and introducing a weighted value for the arrival time: weighted
\begin{equation}
    t_{arrival}=\frac{\sum{A_it_i}}{\sum{A_i}},
\end{equation}
where $A_i$ are the amplitude values produced by the model on each position.

\begin{figure}[H]
\includegraphics[width=0.8\linewidth]{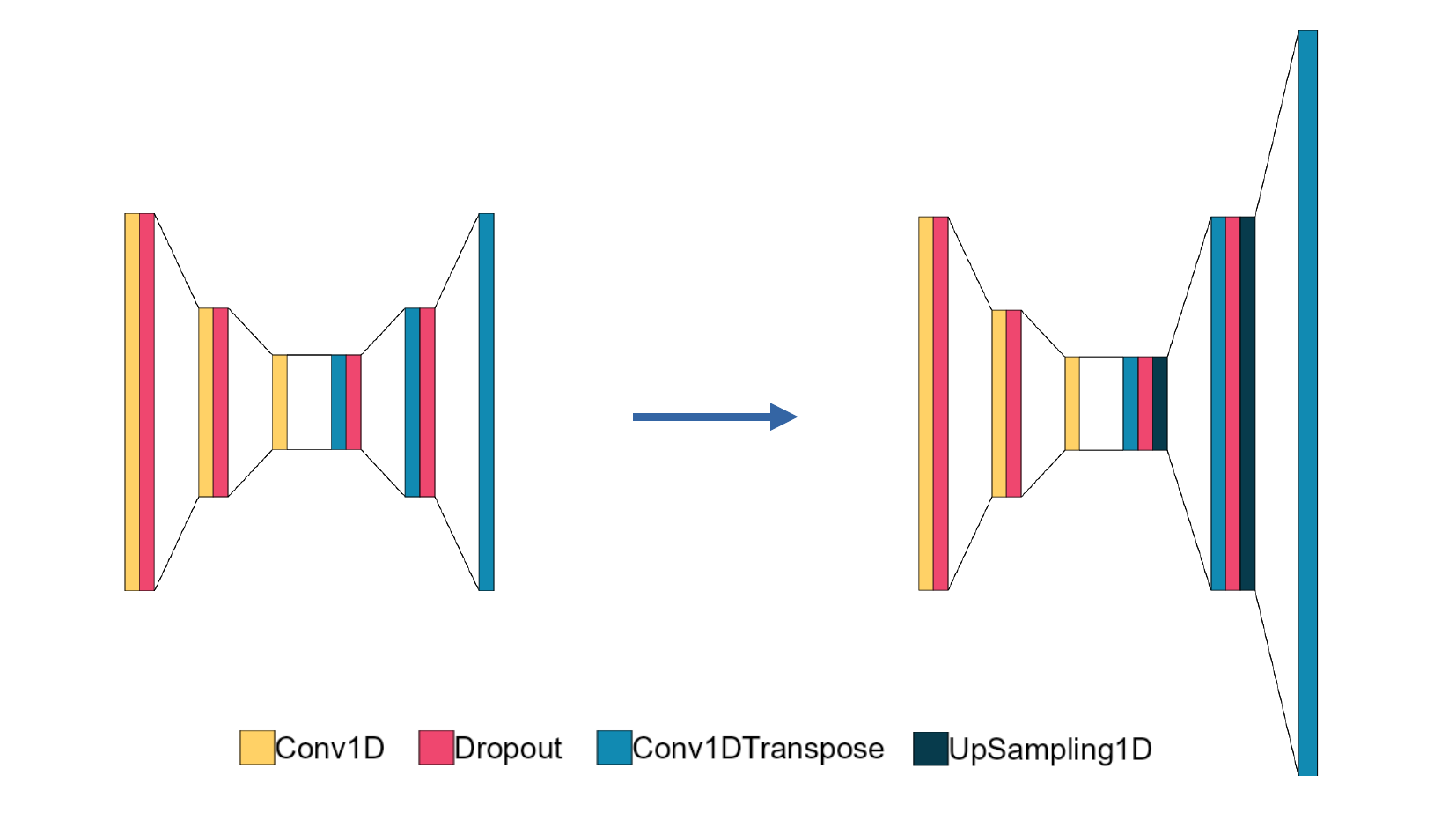}
\caption{Structure of the hidden layers of the MAC and the UMAC. On the left are the hidden layers in the MAC architecture. It consists of an encoder with three convolutional layers with dropout layers between them, followed by a decoder of three transposed convolution layers with dropout layers between them. On the right are the hidden layers of the UMAC architecture: the encoder of the MAC is kept the same and two upsampling layers are placed in the decoder, making the output two times longer after each transposed convolution.
\label{fig:achitectures}}
\end{figure}

In order to probe the time resolution capabilities, 
a new, upsampling version of the used model (UMAC) was developed. It assigns a 4096 values long label vector to each 1024~ns long event. This allows the rounding of the true arrival time value to be carried out with 0.25~ns precision. 
The model architecture was expanded with two upsampling layers with a factor of two in the decoder, each of which doubles the length of the output. They are placed after the first and second transpose convolution layers. The kernel size of the second transposed convolution is adjusted from 12 to 24, and the kernel size of the third transposed convolution is adjusted from 18 to 36. Figure~\ref{fig:achitectures} shows a comparison between the topologies of the hidden layers of the MAC and the UMAC.  The results are post-processed using the weighted arrival time determination.

The MAC and the UMAC performance were compared by applying both models to the same independent dataset, used for the Occlusion Sensitivity investigation. The mean absolute error (MAE) and mean square error (MSE) were evaluated for both the predicted arrival times and signal amplitudes as well as the mean value and standard deviation (SD) of the error distributions. The MAE and MSE are defined as
\begin{equation}
    \text{MAE} = \frac{1}{N}\sum_{i=1}^{N}|\xi^{pred}_i - \xi^{truth}_i|
    \label{eq:mae}
\end{equation}
\begin{equation}
    \text{MSE} = \frac{1}{N}\sum_{i=1}^{N}(\xi^{pred}_i - \xi^{truth}_i)^2,
    \label{eq:mse}
\end{equation}
where $\xi^{pred}_i$ is the predicted value of either the arrival time or the energy,
$\xi^{truth}_i$ is the true value, and N is the number of events in the testing dataset. 
The results can be seen in Table~\ref{table:metrics}. 

\vspace{-9pt}

\begin{table}[H]
    
    \caption{Comparison of the performance metrics of the MAC and the UMAC.   \label{table:metrics}}

    \begin{tabularx}{\textwidth}{Lcc}

         \noalign{\hrule height 1.0pt}
         \diagbox[width=9cm, height=1.1cm]{\textbf{Metrics}}{\textbf{Model}}& \multirow{-1.7}{*}{\textbf{MAC}} & \multirow{-1.65}{*}{\textbf{UMAC}}\\
         
         \noalign{\hrule height 0.5pt}
        Arrival time MAE [ns] & 0.77 & 0.19\\
         \midrule
        Arrival time MSE [ns$^2$] & 0.86 & 0.16 \\
         \midrule
        Amplitude MAE [mV] & 111.6 & 23.2\\
         \midrule
        Amplitude MSE [mV$^2$] & 17,045.8 & 988.6\\
         \midrule
        Arrival time error distribution mean [ns] & 0.75 & 0.02\\
         \midrule
        Amplitude error distribution mean [mV] & $-$111.8 & $-$2.7\\
         \midrule
        Arrival time error distribution SD [ns] & 0.43 & 0.30\\
         \midrule
        Amplitude  error distribution SD [mV] & 67.7 & 31.3\\
        \bottomrule
        
    \end{tabularx}
 
\end{table}

The MAE value for the arrival time, predicted by the UMAC, is more than four times smaller than the MAE for the MAC. The MSE for the UMAC arrival time predictions is more than five times smaller than the MSE for the MAC arrival time predictions. The advantages of the UMAC model are even better pronounced for the amplitude reconstruction where the MAE for the UMAC is close to 5 times smaller than the MAC, and the MSE is more than 17 times smaller.

\begin{figure}[H]
\includegraphics[width=0.75\columnwidth]{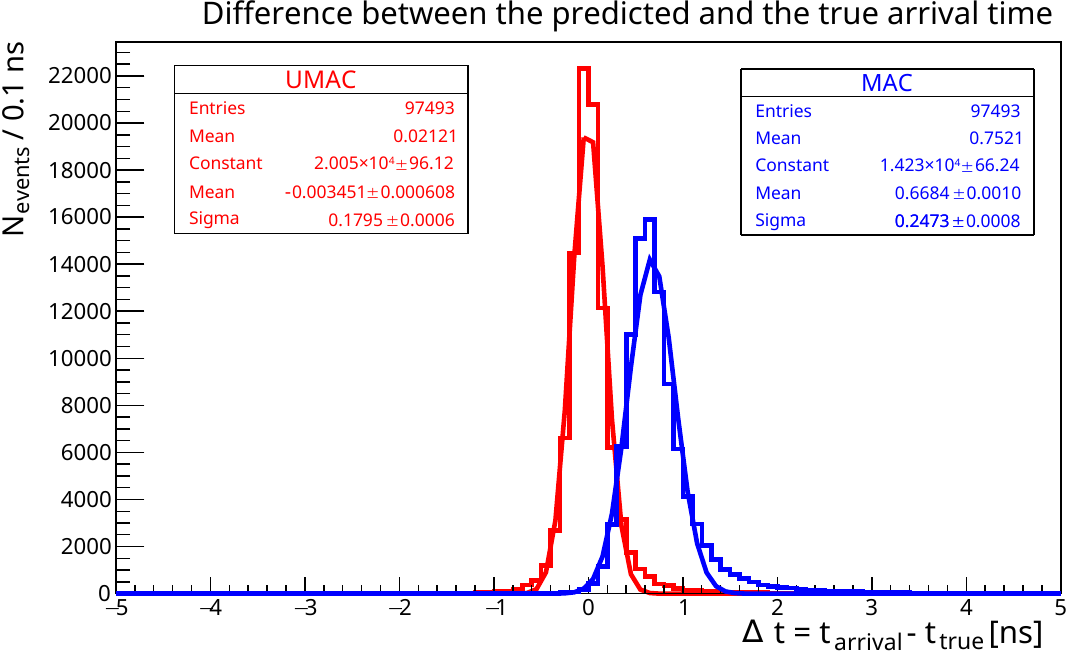}
\caption{
{Difference} 
 {between} 
 the {true} 
 and the predicted arrival time for the MAC (blue) and the UMAC (red). The MAC Gaussian fit shows a mean difference of 0.7~ns, due to the difference between the true signal arrival time and time bin position in the label. The UMAC reduces this difference to $-$0.003~ns and provides a more narrow distribution.\label{fig:up}}
\end{figure}

The difference between the true and the predicted arrival time for the MAC and for the UMAC is shown in Figure~\ref{fig:up}. The two distributions were fit using a Gaussian function. The UMAC results show that most of the events have minimal difference between the predicted and the true arrival time value, compared to the 
0.7~ns mean of the fit for the MAC. The fit 
has a standard deviation of 0.1795 $\pm$ 0.0006~ns for the UMAC, compared to a standard deviation of 0.2473 $\pm$ 0.0008~ns for the MAC.

\begin{figure}[H]
\includegraphics[width=\columnwidth]{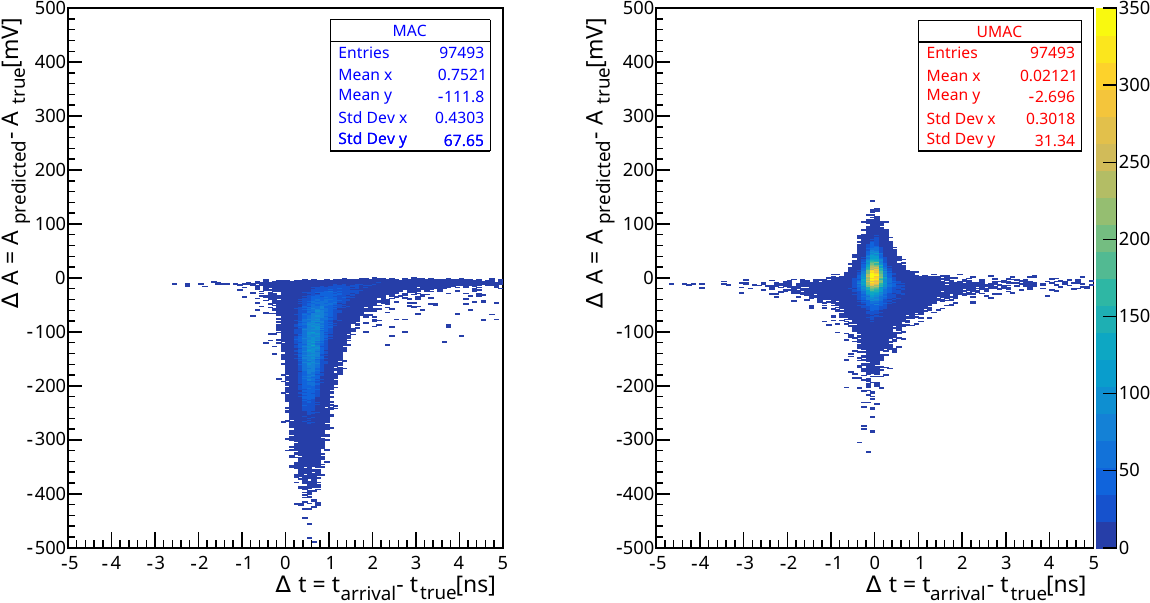}
\caption{{Difference} 
 between the true and the predicted arrival time versus the difference between true and predicted energy for the two models.
(\textbf{Left}) The MAC shows an offset both in the predicted arrival time and the energy from the true values. (\textbf{Right}) The UMAC plot is centered at 0 along both axes, resulting in more accurate reconstruction both in terms of time and amplitude. \label{fig:diff2d}}
\end{figure}

The two-dimensional distribution of the difference between the predicted and true time and the difference between predicted and true amplitude is presented in Figure~\ref{fig:diff2d}.
The MAC shows an offset in the negative values, which means that the reconstructed amplitude is smaller than the true value. This can also be seen in the real data applications of the model, discussed in the previous sections. 
The difference between predicted and true amplitude is well centered at 0 for the UMAC, while for MAC, an additional energy calibration is necessary, 
as pointed out in \cite{bib:calor}.
This suggests that upsampling models do perform better 
for  
reconstructing 
both 
the arrival time and 
for the signal amplitude.

\vspace{-9pt}

\section{Discussion and Conclusions}
Neural networks with autoencoder architectures can be successfully used for signal parameters determination in high-energy physics experiments. Their performance can be further improved by employing explainable AI techniques.
A modified autoencoder model trained on simulated datasets is capable of determining the signal arrival time and amplitude for events with pulses, created by particles in the scintillating crystals of electromagnetic calorimeters. Such models are valuable in cases where significant pile-up is observed, as they show good 
pulse separation abilities.

Performing an Occlusion Sensitivity investigation of the output of the model shows that the signal front and maximum are the most important parts of the pulse shape for its correct identification. Looking into the differences of the Occlusion Sensitivity output for individual events reveals the model's time determination abilities, which surpass the precision of recording the true values in the simulation.

Developing an upsampling version of the model makes it possible to assign a four-times-more precise time label to each 1024~ns long event. One arising disadvantage of such models is that their training and application require more resources and time, as they contain a higher number of parameters. However, the achieved performance in terms of time resolution shows improvement, lowering both the mean offset of the prediction from the true value from 0.7 to 0.003~ns and the standard deviation of distribution from 0.2473 to 0.1795~ns.
In addition, the upsampling models show promising results for the signal amplitude reconstruction which might lead to better energy resolution
without the necessity for subsequent energy calibration
when applying the models to real data.

\vspace{6pt} 

\authorcontributions{Methodology, K.D. and V.K.; Software, K.D. and P.P.; Validation, K.D.; Formal analysis, K.D.; Investigation, P.P.; Writing---original draft, K.D.; Writing---review and editing, V.K. and P.P.; Supervision, V.K. and P.P.; Project administration, V.K. and P.P.; Funding acquisition, V.K. and P.P.
All authors have read and agreed to the published version of the manuscript.} 

\funding{This work is partially supported by BNSF: KP-06-D002\_4/15.12.2020 within MUCCA, CHIST-ERA-19-XAI-009 and by the European Union---NextGenerationEU, through the
National Recovery and Resilience Plan of the Republic of Bulgaria, project SUMMIT BG-RRP-2.004-0008-C01.} 

\dataavailability{The raw data supporting the conclusions of this article will be made available by the authors on request.}



\conflictsofinterest{The authors declare no conflicts of interest. The funders had no role in the design of the study; in the collection, analyses, or interpretation of data; in the writing of the manuscript; or in the decision to publish the results.}


\begin{adjustwidth}{-\extralength}{0cm}
\reftitle{References}



\begin{thebibliography}{999}

\bibitem {bib:3Dclu}
Morgunov, V.;  Raspereza, A. Novel 3-D clustering algorithm and two particle separation with tile HCAL. In Proceedings of  the International Conference, LCWS 2004, {Paris, France, 19--23 April 2004}; pp.  431--436.

\bibitem {bib:PSD}
Shpak, K. {[CALICE]} 
Separation of two electromagnetic or electromagnetic-hadronic showers in CALICE SiW ECAL and ILD.  In Proceedings of the {International Workshop on Future Linear Colliders (LCWS2017), Strasbourg, France, 23--27 October 2017}.



\bibitem {bib:aad}
Aad, G.; {Berthold, A.S.; Calvet, T.; Chiedde, N.; Fortin, E.M.; Fritzsche, N.; Hentges, R.; Olavi Laatu, L.A.; Monnier, E.; Straessner, A.}; et al.  Artificial Neural Networks on FPGAs for Real‑Time Energy
Reconstruction of the ATLAS LAr Calorimeters. \textit{Comput. Softw. Big Sci.} \textbf{2021}, \textit{5}, 19 [\href{http://doi.org/10.1007/s41781-021-00066-y}{CrossRef}]

\bibitem{bib:intgrad}
Sundararajan, M.; {Taly, A.; Yan, Q.} Axiomatic attribution for deep networks. In \textit{ICML'17: Proceedings of the 34th International Conference on Machine Learning, {Sydney, Australia, 6--11 August 2017}}; {PMLR: Birmingham, UK, 2017}; pp. 3319--3328.

\bibitem{Smilkov2017}
Smilkov, D.; Thorat, N.; Kim, B.; Viégas, F.B.; Wattenberg, M. SmoothGrad: Removing noise by adding noise. \textit{arXiv} \textbf{2017}, arXiv:1706.03825.

\bibitem{Valois2023}
Valois, P.; Niinuma, K.; Fukui, K. Occlusion Sensitivity Analysis with Augmentation Subspace Perturbation in Deep Feature Space. In Proceedings of the {2024 IEEE/CVF Winter Conference on Applications of Computer Vision (WACV)}, {Waikoloa, HI, USA, 3--8 January 2024}; pp. 4817--4826.

\bibitem {bib:rojat}
Rojat, T.; {Puget, R.; Filliat, D.; Del Ser, J.; Gelin, R.; Díaz-Rodríguez, N.} Explainable Artificial Intelligence (XAI) on Timeseries Data: A Survey. \textit{arXiv} \textbf{2021}, arXiv:2104.00950.

\bibitem {bib:tf}
Abadi, M.; Agarwal, A.; Barham, P.; Brevdo, E.; Chen, Z.; Citro, C.; Corrado, G.S.; Davis, A.; Dean, J.; Devin, M.; {et al.} 
TensorFlow: Large-Scale Machine Learning on Heterogeneous Systems. {2015}. Available online: \url{https://tensorflow.org} ({accessed on 22 January~2025}).

\bibitem {bib:keras}
Chollet, F. Keras. {2015}. {Available} 
online: \url{https://keras.io} ({accessed on 22 January 2025}).


\bibitem {bib:matplotlib}
Hunter, J.D. Matplotlib: A 2D Graphics Environment. {\it{Comput. Sci. Eng.}} \textbf{2007}, \textit{9}, 90--95. [\href{http://dx.doi.org/10.1109/MCSE.2007.55}{CrossRef}]

\bibitem {bib:ROOT}
Brun, R.; Rademakers, F.
\newblock {ROOT: An object oriented data analysis framework}.
\newblock {\em Nucl. Instrum. Methods Phys. Res. Sect. A Accel. Spectrometers Detect. Assoc. Equip.} {\bf 1997}, {\em 389}, 81--86. [\href{http://dx.doi.org/10.1016/S0168-9002(97)00048-X}{CrossRef}]

\bibitem {bib:viskeras}
Gavrikov, P. Visualkeras. {2020}. Available online: \url{https://github.com/paulgavrikov/visualkeras} ({accessed on 22 January 2025}).

\bibitem{bib:calor}
Dimitrova, K.; {PADME Collaboration}. Using Artificial Intelligence in the Reconstruction of Signals from the PADME Electromagnetic Calorimeter. \textit{Instruments} \textbf{2022}, \textit{6}, 46. [\href{http://dx.doi.org/10.3390/instruments6040046}{CrossRef}]

\bibitem{bib:nafski22}
Buchakchiev, V.; {Dimitrova, K.; Georgiev, G.; Georgieva, G.;  Kozhuharov, V.}  Pattern recognition and signal parameters extraction using machine learning methods. \textit{J. Phys. Conf. Ser.} \textbf{2023}\textit{, 2668,} 012001. [\href{http://dx.doi.org/10.1088/1742-6596/2668/1/012001}{CrossRef}]

\bibitem{bib:azuara}
Azuara de Pablo, G.; {Barrera López de Turiso, E.; Ruiz González, M.; Barba Navarrete, P.L.; Piñas Higueruela, A.}   Convolutional Autoencoders for Signal Reconstruction and their Application to Damage Signature Extraction. In {Proceedings of the 10th European Workshop on Structural Health Monitoring (EWSHM 2024), Potsdam, Germany, 10--13 June 2024}; {Volume 29,} p. 7.



\bibitem{bib:caen}CAEN 2012 {{V1742 Technical Information Manual Rev. 6; V1751 Technical Information Manual Rev. 12}}. Available online: \url{http://www.caen.it} ({accessed on 22 January 2025}).

\bibitem{bib:nafski23}
Dimitrova, K.; {PADME Collaboration}. Machine learning assisted reconstruction of positron-on-target annihilation events in the PADME experiment. \textit{J. Phys. Conf. Ser.} \textbf{2024}, \textit{2794}, 012001. [\href{http://dx.doi.org/10.1088/1742-6596/2794/1/012001}{CrossRef}]

\bibitem{bib:varna}
Dimitrova, K.; Kozhuharov, V.; Petkov, P. 
Applicability evaluation of selected xAI methods for machine learning algorithms for signal parameters extraction. In Proceedings of the {Second Workshop on Soliton Theory, Nonlinear Dynamics and Machine Learning, Varna, Bulgaria, 16--21 August 2024}.

\bibitem {bib:occ}
Zeiler, M.D.; Fergus, R. Visualizing and Understanding Convolutional Networks. In \textit{Computer Vision---ECCV 2014, Proceedings of the 13th European Conference, Zurich, Switzerland, 6--12 September 2014}; {Springer}:  {Berlin/Heidelberg, Germany,} 2014; pp. 818--833. 




\end{thebibliography}


\PublishersNote{}
\end{adjustwidth}
\end{document}